# Creation and characterization of He-related color centers in diamond

J. Forneris[1,2,3,*], A. Tengattini[2,1,3], S. Ditalia Tchernij[2,1,3], F. Picollo[1,2,3], A. Battiato[2,1,3], P. Traina[4], I. P. Degiovanni[4], E. Moreva[4], G. Brida[4], V. Grilj[5], N. Skukan[5], M. Jakšić[5], M. Genovese[4,1,3], P. Olivero[2,1,3]

[1]*Istituto Nazionale di Fisica Nucleare (INFN), Sez. Torino, via P. Giuria 1, 10125, Torino, Italy*
[2]*Physics Department and "NIS " Inter-departmental Centre - University of Torino; via P. Giuria 1, 10125, Torino, Italy*
[3]*Consorzio Nazionale Interuniversitario per le Scienze Fisiche della Materia (CNISM), Sez. Torino, Torino, Italy*
[4]*Istituto Nazionale di Ricerca Metrologica (INRiM), Strada delle Cacce 91, 10135 Torino, Italy*
[5]*Ruđer Bošković Institute, Bijenicka 54, P.O. Box 180, 10002 Zagreb, Croatia*

*forneris@to.infn.it

## Abstract

Diamond is a promising material for the development of emerging applications in quantum optics, quantum information and quantum sensing. The fabrication and characterization of novel luminescent defects with suitable opto-physical properties is therefore of primary importance for further advances in these research fields.

In this work we report on the investigation in the formation of photoluminescent (PL) defects upon MeV He implantation in diamond. Such color centers, previously reported only in electroluminescence and cathodoluminescence regime, exhibited two sharp emission lines at 536.5 nm and 560.5 nm, without significant phonon sidebands.

A strong correlation between the PL intensities of the above-mentioned emission lines and the He implantation fluence was found in the $10^{15}$-$10^{17}$ cm$^{-2}$ fluence range. The PL emission features were not detected in control samples, i.e. samples that were either unirradiated or irradiated with different ion species (H, C). Moreover, the PL emission lines disappeared in samples that were He-implanted above the graphitization threshold. Therefore, the PL features are attributed to optically active defects in the diamond matrix associated with He impurities. The intensity of the 536.5 nm and 560.5 nm emission lines was investigated as a function of the annealing temperature of the diamond substrate. The emission was observed upon annealing at temperatures higher than 500 °C, at the expenses of the concurrently decreasing neutral-vacancy-related GR1 emission intensity. Therefore, our findings indicate that the luminescence originates from the formation of a stable lattice defect. Finally, the emission was investigated under different laser excitations wavelengths (i.e. 532 nm and 405 nm) with the purpose of gaining a preliminary insight about the position of the related levels in the energy gap of diamond.

## 1. Introduction

In the last decade diamond has been thoroughly investigated as a promising material for applications in the field of quantum optics, due to the discovery and the characterization of several luminescent point defects with appealing light emission and spin properties [1,2, 3].

Up to date, the most prominent system in this field is the negatively-charged nitrogen-vacancy center (NV$^-$), whose well established high quantum efficiency, stability at room temperature and appealing spin properties structure represent an intriguing potential for applications in quantum photonics, cryptography, sensing and computing [4-9]. On the other hand, some of its limitations (relatively long radiative lifetime, charge state blinking and broad spectral emission [10]) led to the exploration of alternative luminescent centers for single-photon source applications, such as the Si-V center [11,12], Ni- [13-15], Eu- [16] and Ge-related [17,18] impurities, as well as radiation-damage related defects [19] and other bright NIR emitters [20,21], which demonstrated up to tenfold higher emission rates, as well as a strongly polarized and spectrally narrower emission. Thus, the fabrication of novel luminescent defects with desirable properties upon the implantation of selected ion species still represents a crucial strategy to achieve further advances in the aforementioned research fields.

In this work, we report on the investigation of the photoluminescence (PL) properties of optically active defects in diamond obtained upon the implantation of MeV He$^+$ ions and subsequent thermal annealing at temperatures >500 °C. The measured PL emission consists of two sharp and intense emission lines at 536.5 nm and 560.5 nm, characterized by a negligible phonon coupling. These emission lines were reported, together with a series of other minor PL peaks, in two previous cathodoluminescence studies on He-implanted IIa-type diamond [22,23], and more recently were identified in electroluminescence spectra acquired from He-implanted diamond devices [24]. Up to date, the excitation of these emission lines was never reported in PL regime, despite the wider availability of this characterization technique. This somewhat surprising evidence is motivated by the spectral range of this emission, which is typically filtered out in the most commonly adopted PL confocal microscopes, as it is comprised between the excitation wavelength (typically 532 nm) and the first-order Raman line (at 572.5 nm for the above-mentioned excitation). Differently from the commonly adopted experimental PL setups, in the present work two alternative laser excitation wavelengths (i.e., 532 nm in combination with narrowband notch filtering, and 405 nm) were adopted.

## 2. Experimental

The experiments were performed on a set of type-IIa single-crystal diamond samples: 3 nominally identical "optical grade" 3×3×0.3 mm$^3$ substrates by ElementSix (samples #1-3), with respectively <1 ppm and <0.05 ppm concentrations of substitutional nitrogen and boron, and a 2×2×0.3 mm$^3$ "detector grade" ElementSix substrate (sample #4) with <5 ppb nitrogen and boron concentrations. He implantations were performed at the AN2000 microbeam line of the INFN National Laboratories of Legnaro. Several 300×300 μm$^2$ regions were homogeneously irradiated on sample #1 with a 1.3 MeV He$^+$ raster scanning micro-beam at fluences of 1×10$^{16}$ cm$^{-2}$, 2×10$^{16}$ cm$^{-2}$ and 1×10$^{17}$ cm$^{-2}$. An additional 100×100 μm$^2$ irradiation was performed on sample #2 with a 1 MeV He$^+$ beam at a 1×10$^{15}$ cm$^{-2}$ fluence. A 200×200 μm$^2$ implantation at 5×10$^{14}$ cm$^{-2}$ fluence was performed on sample #4 with a 1.8 MeV He$^+$ beam. Sample #3 underwent a control implantation with 6 MeV C$^+$ ions at 2×10$^{15}$ cm$^{-2}$ fluence, at the Laboratory for Ion Beam Interactions of the Ruđer Bošković Institute (Zagreb). The implantation fluence was chosen to obtain a similar vacancy density in the diamond substrate in correspondence of the Bragg peak (i.e. 1×10$^{22}$ vacancies cm$^{-3}$) as that achieved in the case of 1.8 MeV He$^+$ implantation at 1.5×10$^{16}$ cm$^{-2}$ fluence [25]. After ion implantation, samples #1–3 were annealed in vacuum at 1000 °C for 2 hours to promote the formation and optical activation of the luminescent defects. Sample #4 underwent subsequent annealing steps in vacuum (2 hours each) at increasing temperatures up to 1000 °C, to investigate the role of thermal processing in the defect formation. After each thermal treatment, the ion-implanted region of sample #4 was characterized in PL emission. PL spectra were acquired with a Horiba Jobin Yvon HR800 Raman micro-spectrometer equipped with a 1800 mm$^{-1}$ diffraction grating. The optical excitation was provided by a continuous 532 nm laser focused with a 20× air objective. The excitation radiation was filtered out from the CCD detection system by a narrow-band notch filter (Super Notch Plus 532 nm filter, 6.0 optical density, <10 nm spectral bandwidth).

PL measurements at a lower excitation wavelength were performed using a confocal microscope at the Italian National Institute for Metrological Research (INRiM). In this case, the optical excitation was provided by a continuous 405 nm laser focused with a 100× air objective. The PL signal was then acquired by a Si-single-photon-avalanche photo-diode (SPAD) operating in Geiger mode. The use of a dichroic mirror (Semrock 442 nm laser BrightLine) and the optical filters (Thorlabs FEL0500) allowed spectral measurements at wavelengths larger than 500 nm. The PL spectra were acquired using a single-grating monochromator (1600 grooves mm$^{-1}$, 600 nm blaze) connected to the afore-mentioned SPAD.

## 3. Results and discussion

*3.1 Photoluminescence features of He-implanted diamond*

Photoluminescence measurements with 532 nm excitation were acquired in the 533-800 nm spectral range from the regions implanted at increasing ion fluences of samples #1 and #2. In order to provide a quantitative comparison between the different emission intensities, the PL spectra were normalized to the first-order diamond Raman peak measured at 572.5 nm. The resulting spectra are shown in **Fig. 1**, together with a control PL spectrum acquired from an unimplanted region of sample #1.

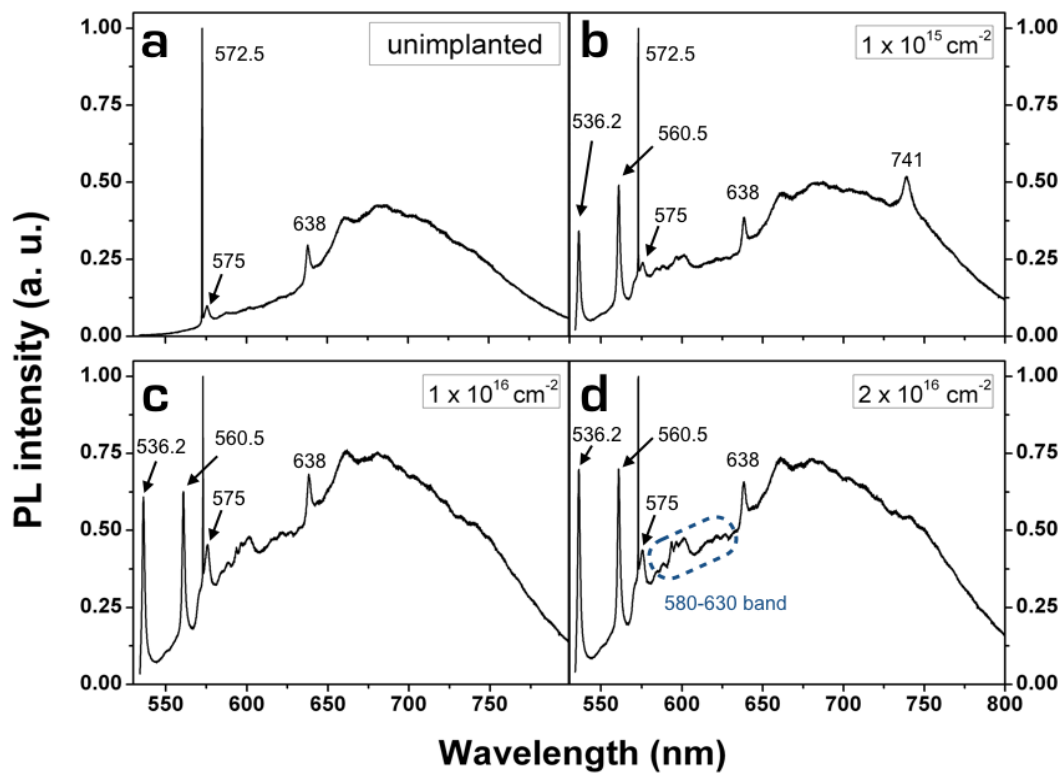

**Figure 1:** PL spectra acquired at 532 nm excitation from He-irradiated "optical grade" diamond substrates: **a)** control unirradiated region (sample #1); **b)** 1 MeV He$^+$ irradiation, 1×10$^{15}$ cm$^{-2}$ fluence (sample #2); **c)** 1.3 MeV He$^+$ irradiation, 1×10$^{16}$ cm$^{-2}$ fluence (sample #1); **d)** 1.3 MeV He$^+$ irradiation, 2×10$^{16}$ cm$^{-2}$ fluence (sample #1).

The PL spectrum of the unimplanted sample (**Fig. 1a**) exhibits the typical features of an "optical grade" diamond substrate: the intense first-order Raman peak at 572.5 nm (i.e. 1332 cm$^{-1}$ Raman shift), the zero-phonon lines (ZPLs) of the NV$^0$ and NV$^-$ centers respectively at 575 nm and 638 nm, together with the corresponding phonon replica at higher wavelengths [4]. Apart from the well-established increase in NV emission intensity, the PL spectra from the implanted regions (**Fig. 1b-1d**) revealed two sharp (<2 nm FWHM) and intense peaks at 536.5 nm and 560.5 nm, together with an emission band in the 580-630 nm spectral range (highlighted in blue in **Fig. 1d** for ease of view). This latter feature cannot be attributed to Raman transitions (as confirmed with subsequent PL measurements with 405 nm excitation, see below), and it is unlikely to be related to either phonon replica of the NV$^0$ and NV$^-$ centers, due to the high intensity contrast with respect to the baseline of the NV$^0$ emission spectrum at wavelengths >610 nm. More likely, it is to be attributed to the phonon sideband of the 560.5 nm transition, consistently with what reported in Ref. [23]. The intensity of these spectral features displayed a clear correlation with the implantation fluence (**Fig. 2**). Additionally, a weak peak at 740 nm, observed at the lowest implantation fluence, is attributed to residual radiation-induced vacancies (GR1 centers), i.e. to isolated vacancies which were not completely annealed out upon thermal treatment [22, 24].

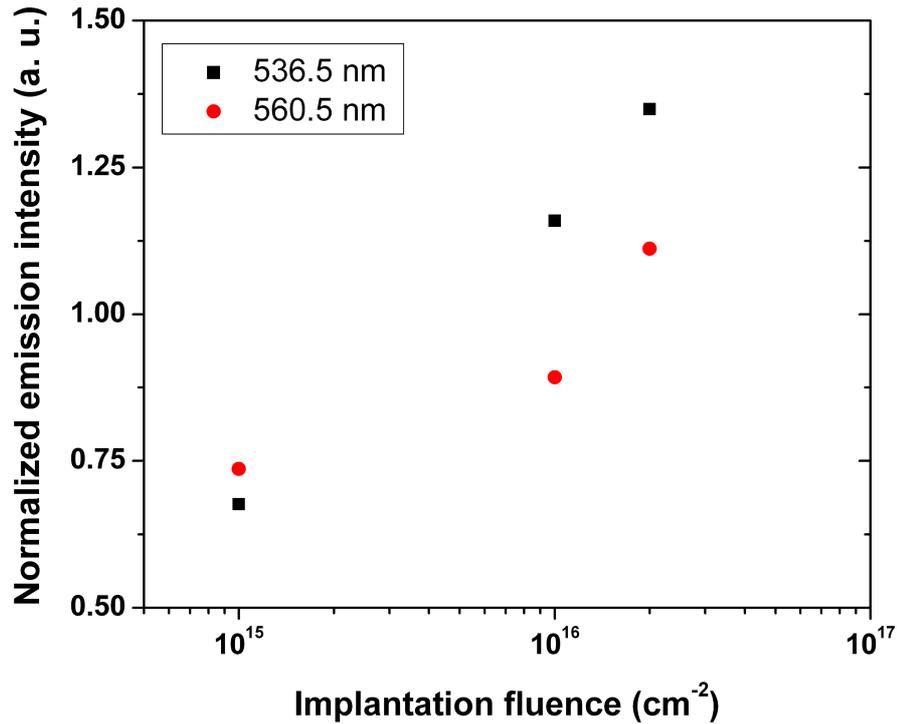

**Figure 2:** Integrated intensity of the 536.5 nm and 560.5 nm PL emission lines reported in Fig. 1 as a function of the implantation fluence.

The PL spectrum is similar to those reported in previous works in both CL [22, 23] and EL [24] regimes. In addition, the complex series of emission lines observed in the 536-575 nm range reported in Ref. [23] was not observed, neither in the present experiment nor in previous EL investigations of He implanted diamond [24]. This finding is consistent with the reported absence of such emission lines in diamond substrates with low nitrogen content [22], as in the case of the samples under investigation.

It is worth noting that the 536.5 nm and 560.5 nm luminescence does not correspond to any emission line associated with the atomic He spectrum [26]. The origin of the PL emission should then be either attributed to the formation of lattice defects, whose abundance correlates with the He implantation fluence [23], or to the modification of the above-mentioned atomic He emission spectrum by surrounding radiation-induced defects [22].

To address this latter point, a PL spectrum (**Fig. 3a**) was acquired from the region of sample #1 irradiated with 1.3 MeV He$^+$ ions at a fluence of $1\times10^{17}$ cm$^{-2}$, i.e. close to the graphitization threshold of diamond [24]. In this case, the spectrum was not normalized to the intensity of the first-order Raman peak, which was drastically reduced by ion-induced damage and shifted to a position of 573.0 nm (i.e. 1345 cm$^{-1}$ Raman shift). The attribution of the emission at 593.5 nm is unclear, and it can be tentatively interpreted as a PL feature related to ion-induced structural damage, i.e. the 2.087 eV center in diamond [22, 27]. Moreover, the NV$^0$ and NV$^-$ ZPLs are not visible. The PL spectrum is instead dominated by a large emission in the 740-800 nm range, which is typically associated with the formation of the radiation B-band [22] associated to lattice dislocations. The whole spectrum shows a periodic intensity modulation which is attributed to interference fringes caused by multiple internal reflections of the broad luminescence emission occurring between the sample surface and the buried damaged layer [28]. Most remarkably, the absence of the PL spectral lines at 536.5 nm and 560.5 nm indicates that these emission lines must be associated with defects in the diamond lattice and that they cannot be related to the presence of a graphitic/amorphous phase. Moreover, their absence also suggests that the afore-mentioned lines do not originate from a modification of the atomic He spectrum, as a detectable emission would still be expected considering the large amount of the He impurities introduced in the diamond lattice at this given implantation fluence. The two emission lines at 536.5 nm and 560.5 nm are thus attributed to the formation of defects in the diamond lattice, whose abundance correlates with the He implantation fluence [23] as long as the diamond structure is not subjected to amorphization.

With the purpose of further strengthening the attribution of the PL emissions at 536.5 nm and 560.5 nm to He-related defects, additional PL measurements were acquired from sample #3, after 6 MeV $C^+$ implantation and thermal annealing. The emission spectrum (**Fig. 3b**) clearly exhibits the typical features of $NV^0$ and $NV^-$ emissions, together with a residual component of GR1 centers at 740 nm. Also in this case, the emission lines at 536.5 nm and 560.5 nm could not be detected within the experimental sensitivity. Moreso, it is worth mentioning that the emission band in the 580-630 nm region was not observed in neither the high-fluence He implanted sample, nor in the C implanted sample, thus preventing its attribution as a generic radiation-induced defect, in agreement with the attribution to a phonon sideband of the 560.5 nm peak [23].

Even if the reported spectra do not allow a fully conclusive attribution, which can only be provided with stress-dependent PL measurements to study the symmetry of the defect, to be correlated with a suitable theoretical model, nonetheless they provide a strong indication that the 536.5 nm and the 560.5 nm spectral features (*He-related centers* in the following) are not related to intrinsic defects, interstitials, or generic radiation effects due to light ions implantation in diamond. Rather, they indicate that the presence of He atoms plays a key role in the formation of a specific optical center in two possible charge states, or alternatively of two types of He-related complexes.

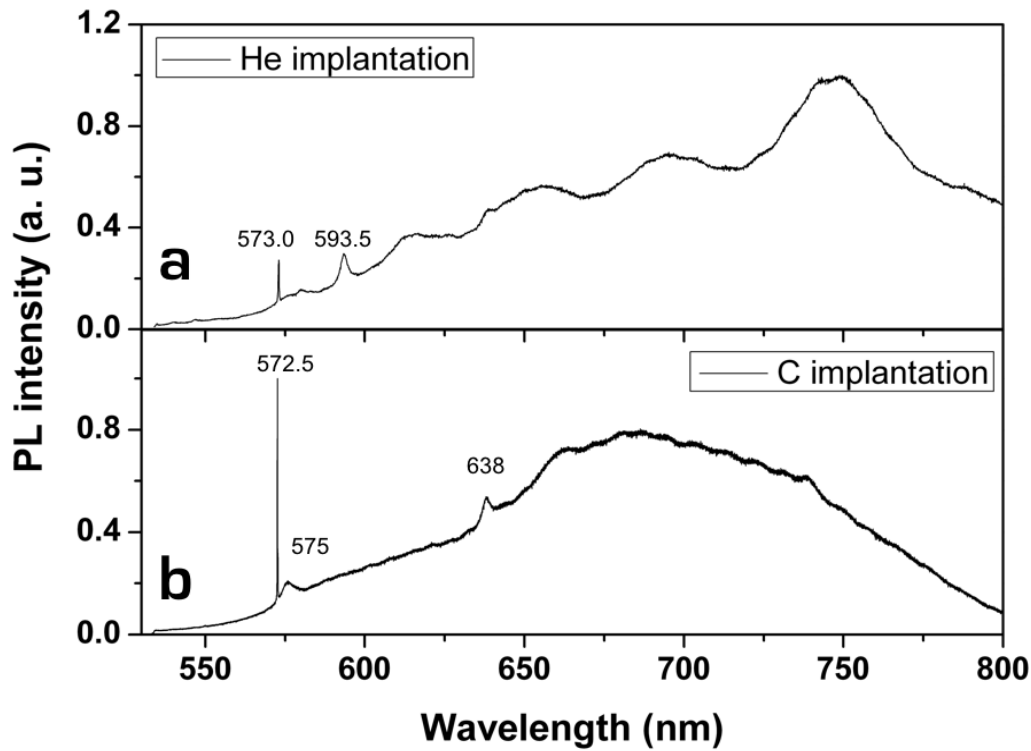

**Figure 3:** PL spectra acquired under 532 nm excitation from **a)** sample #1 irradiated with a 1.3 MeV He$^+$ beam at $1\times10^{17}$ cm$^{-2}$ fluence **b)** sample #3 irradiated with a 6 MeV C$^+$ beam at $2\times10^{15}$ cm$^{-2}$ fluence.

*3.2 Thermal annealing*

The intensity of the emission lines from the He-related centers was studied on sample #4 as a function of the annealing temperature, to investigate the role of the thermal processing in the formation of the defect. The PL spectrum acquired from sample #4 in its pristine conditions (**Fig. 4a**) only displays the first- and second-order Raman features, while no significant PL features are visible over the whole spectral range, as expected from the high purity of this sample. The PL spectrum acquired from the as-implanted sample (**Fig. 4b**) reveals an intense emission band in the 720-800 nm range, dominated by the GR1 peak at 741 nm and associated with the introduction of radiation damage in the diamond lattice. Compatibly with what reported in previous studies [22], the intensity of the GR1 emission does not significantly change upon 250 °C (**Fig. 4c**) and 500 °C (**Fig. 4d**) annealing, while it dramatically decreases after processing at higher temperatures (i.e., 750 °C, **Fig. 4e**). At even higher temperatures (i.e., 1000 °C, **Fig. 4f**), it completely disappears.

On the other hand, the He-related PL lines appear concurrently with the disappearance of the GR1 feature, and increase in their intensity at 1000 °C annealing. This evidence seems to confirm that the 536.5 nm and 560.5 nm spectral lines are associated with the formation of specific complex(es), as proposed originally in Ref. [23]. It is worth remarking that, regardless of the attribution of the two lines to two different He-related defects or rather to one defect in two possible charge states, these complexes display a higher structural stability with respect to the isolated vacancy defect. On the other hand, the absence of NV emission features (with the possible exception of a very weak NV$^0$ emission feature upon 1000 °C annealing) indicates that nitrogen is unlikely to play any role in the formation of this complex, as previously inferred from the CL characterization of He-implanted diamond [22].

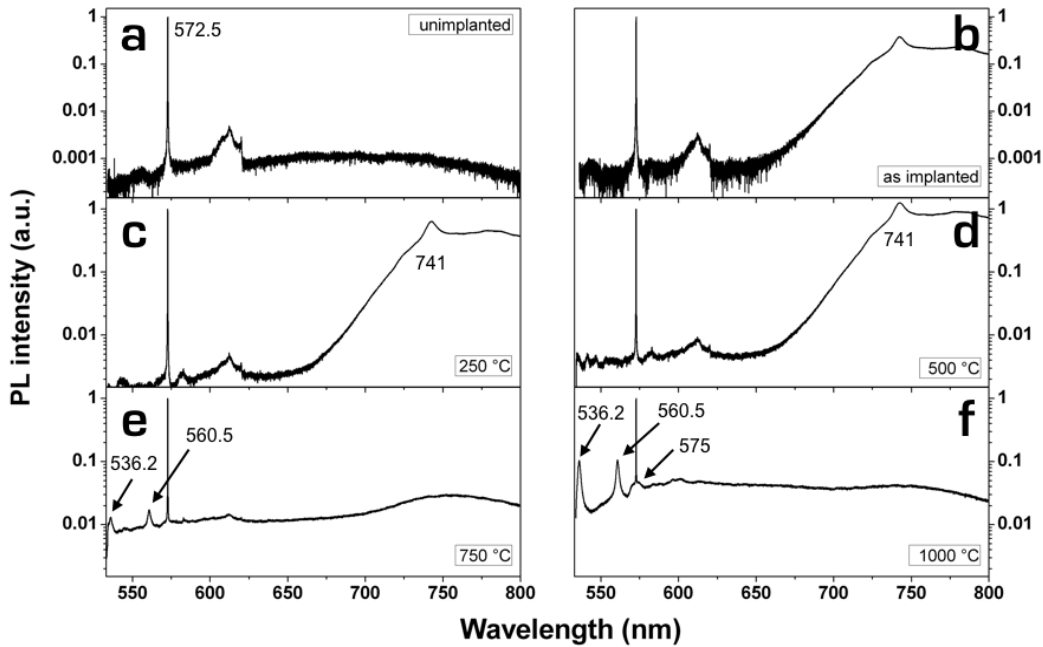

**Figure 4:** PL spectra acquired under 532 nm excitation from sample #4 upon 2-hours annealing steps at increasing temperatures: **a)** pristine sample; **b)** after implantation with 1.8 MeV He$^+$ ions at 5×10$^{14}$ cm$^{-2}$ fluence; **c)** after 250 °C annealing; **d)** after 500 °C annealing; **e)** after 750 °C annealing; **f)** after 1000 °C annealing.

*3.3 Photoluminescence at 405 nm excitation*

The PL emission from the region of sample #1 implanted with 1.3 MeV He$^+$ ions at a a 1×10$^{16}$ cm$^{-2}$ fluence was investigated under 405 nm excitation. The PL spectrum (**Fig. 5a**) is compared with the corresponding PL spectrum acquired under 532 nm excitation (**Fig. 5b**) in the same experimental conditions. The two PL spectra could not be normalized to the first-order Raman line, which is filtered out by the optical filters under blue laser stimulation. Instead, a normalization to the intensity of the 560.5 nm PL peak was chosen for ease of visualization.

Both the 536.5 nm and 560.5 nm emission lines are clearly visibile under 405 nm excitation, indicating that this wavelength is effective at optically stimulating the photoluminescence of He-related defects. As mentioned before, it is worth noting that, similarly to what observed under 532 nm excitation, the PL band centered at 600 nm is still detected, thus ruling out its possible attribution as a Raman feature. The emission is thus identified as an additional spectral feature of He-related defects in diamond, and it is tentatively attributed to a phonon sideband of the He-related peak at 560.5 nm.

The observation of He-related PL features under 405 nm excitation provides some useful indications on the position of the corresponding levels in the energy band gap of diamond, while unequivocally ruling out their attributions to Raman transitions. In more details, our results indicate that the ground state(s) of the He-related center(s) lie at an energy of >3.06 eV from the bottom of the conduction band, and can therefore be regarded as very deep states in the forbidden gap. Furthermore, it is worth noting that no emission lines at wavelengths shorter than 536.5 nm were observed, such as the intense and sharp peak at 522.5 nm reported in Ref. [23], which was correlated with the 536.5 nm and 560.5 nm transitions. In this regard, our result seems to counter-proof this tentative attribution of the 522.5 nm feature. In principle, the absence of this spectral feature could be explained by the fact that it cannot be excited under 532 nm radiation; on the other hand, its absence under 405 nm laser excitation could indicate that the ground state of such defect is placed at less than 3.06 eV from the bottom of the conduction band. It is however worth noting that the 522.5 nm emission was not observed under electroluminescence regime in a recent experiment on a type-IIa diamond substrate [24]. This could indicate that the afore-mentioned spectral lines should be interpreted as a feature of the specific diamond substrate considered in Ref. [23].

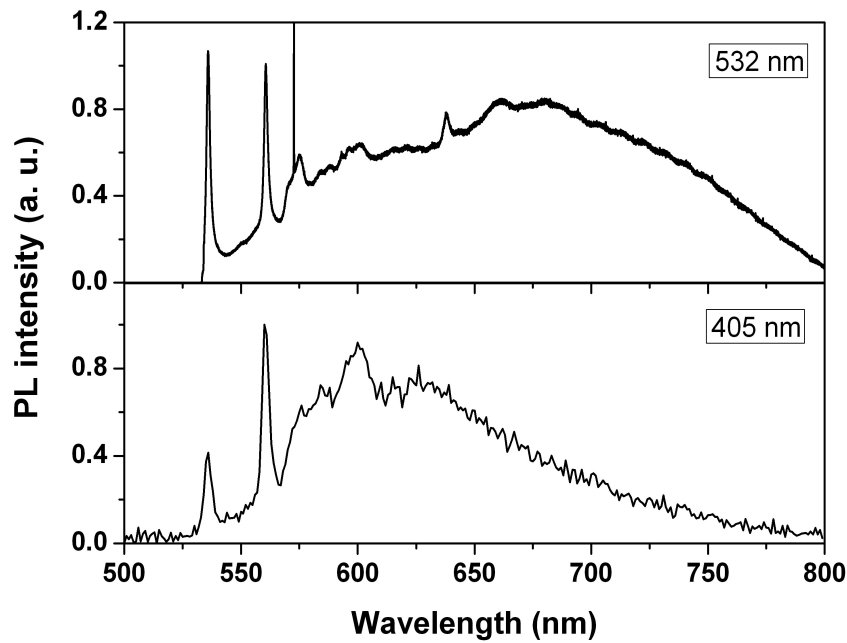

**Figure 5:** PL spectra acquired under different excitation wavelengths from sample #1, upon the 1.3 MeV He$^+$ irradiation at $1\times10^{16}$ cm$^{-2}$ fluence and subsequent thermal annealing: **a)** 532 nm excitation; **b)** 405 nm excitation.

## 4. Conclusions

In this work we reported for the first time the photo-excitation of two sharp emission lines at 536.5 nm e 560.5 nm in He-implanted diamond. On the basis of the correlation of the PL intensity with the implantation fluence, as well as with the thermal annealing temperature of the implanted substrates, we attributed these spectral features to complexes containing both He impurities and (most likely) vacancies. Moreover, by carrying PL measurements under 405 nm laser excitation, we estimated a lower limit of 3.06 eV on the difference in energy between the ground state of these optical transitions and the bottom of the conduction band.

The results presented in this work provide an interesting perspective on the fabrication of diamond color centers with appealing emission features (sharp emission, low phonon coupling), by means of implantation of largely accessible and easily focusable ions such as He$^+$ [29]. In future works we envisage the characterization of the opto-physical properties (emission lifetime, polarization, spin rensonances) of these centers at the single-defect level, in view of possible applications in quantum optics.

## Acknowledgements

This research activity was supported by the following projects: "DIESIS" project funded by the Italian National Institute of Nuclear Physics (INFN) - CSN5 within the "Young research grant" scheme; "FIRB Future in Research 2010" project (CUP code: D11J11000450001) funded by the Italian Ministry for Teaching, University and Research (MIUR); "A.Di.N-Tech." (CUP code: D15E13000130003) funded by the University of Torino and Compagnia di San Paolo within the "Progetti di ricerca di Ateneo 2012"; "Compagnia di San Paolo" project "Beyond classical limits in measurements by means of quantum correlations"; EMRP Project No. EXL02-SIQUTE; EMPIR Project. No. 14IND05-MIQC2. MeV He implantations were performed within the "Dia.Fab." experiment at the INFN-LNL laboratories.